\documentclass[aps,amsmath, prb,twocolumn,superscriptaddress]{revtex4-2}
\usepackage{lineno}
\usepackage{graphicx}
\usepackage{dcolumn}
\usepackage{bm}
\usepackage{lipsum}
\usepackage{color}
\usepackage{amsfonts,eucal,mathrsfs,amssymb}
\DeclareMathAlphabet{\mathscrbf}{OMS}{mdugm}{b}{n}

\usepackage[table]{xcolor}
\graphicspath{{./figs/}}
\usepackage{multirow}
\usepackage[breaklinks=true,unicode=true,urlcolor = blue,colorlinks = true,citecolor = blue,linkcolor = blue]{hyperref}

\renewcommand{\vec}[1]{\bm{#1}}
\newcommand{\dd}{\mathrm{d}}

\begin{document}

\preprint{}
%
%
\title{Chiral magnetic excitations and domain textures of $g$-wave altermagnets}
%
\author{Volodymyr P. Kravchuk}
\email{v.kravchuk@ifw-dresden.de}
\affiliation{Institute for Theoretical Solid State Physics, Leibniz Institute for Solid State and Materials Research Dresden, D-01069 Dresden, Germany}
\affiliation{Bogolyubov Institute for Theoretical Physics of the National Academy of Sciences of Ukraine, 03143 Kyiv, Ukraine}

\author{Kostiantyn V. Yershov}
\affiliation{Institute for Theoretical Solid State Physics, Leibniz Institute for Solid State and Materials Research Dresden, D-01069 Dresden, Germany}
\affiliation{Bogolyubov Institute for Theoretical Physics of the National Academy of Sciences of Ukraine, 03143 Kyiv, Ukraine}

\author{Jorge I. Facio}
\affiliation{Centro Atomico Bariloche, Instituto de Nanociencia y Nanotecnologia~(CNEA-CONICET) and Instituto Balseiro, Av. Bustillo, 9500, Argentina}

\author{Yaqian~Guo}
\affiliation{Institute for Theoretical Solid State Physics, Leibniz Institute for Solid State and Materials Research Dresden, D-01069 Dresden, Germany}

\author{Oleg~Janson}
\affiliation{Institute for Theoretical Solid State Physics, Leibniz Institute for Solid State and Materials Research Dresden, D-01069 Dresden, Germany}

\author{Olena~Gomonay}
\affiliation{Institut f\"{u}r Physik, Johannes Gutenberg-Universit\"{a}t Mainz, Staudingerweg 7, D-55099 Mainz, Germany}

\author{Jairo Sinova}
\affiliation{Institut f\"{u}r Physik, Johannes Gutenberg-Universit\"{a}t Mainz, Staudingerweg 7, D-55099 Mainz, Germany}
\affiliation{Department of Physics, Texas A\&M University, College Station, Texas 77843-4242, USA}

\author{Jeroen van den Brink}
\affiliation{Institute for Theoretical Solid State Physics, Leibniz Institute for Solid State and Materials Research Dresden, D-01069 Dresden, Germany}
\affiliation{Institute of Theoretical Physics and W{\"u}rzburg-Dresden  Cluster of Excellence {\it ct.qmat}, Technische Universit{\"a}t Dresden, 01062 Dresden, Germany} 

\begin{abstract}
Altermagnets (AMs) constitute a novel class of spin-compensated materials in which opposite-spin sublattices are connected by a crystal rotation, causing their electronic iso-energy surfaces to be spin-split. 
While cubic and tetragonal crystal symmetries tend to produce AMs in which the splitting of {\it electronic} iso-energy surfaces has $d$-wave symmetry, hexagonal AMs, such as CrSb and MnTe, are $g$-wave AMs.
%
%
Here we investigate the purely {\it magnetic} modes and spin-textures of $g$-wave AMs and show that they are drastically different for easy-axial (CrSb) and easy-planar (MnTe) materials. 
We show that in CrSb the splitting of the chiral magnon branches possesses $g$-wave symmetry, with each branch carrying a fixed momentum-independent magnetic moment. The altermagnetic splitting is not affected
by the easy-axial anisotropy and is the same as that in the nonrelativistic limit. 
The magnon splitting of MnTe, however, does not strictly possess $g$-wave symmetry due to its easy-planar anisotropy. Instead the magnetic moment of each branch becomes momentum-dependent, with a distribution that is of $g$-wave symmetry. 
To generalize the concept of the altermagnetic splitting beyond the nonrelativistic limit, we introduce alternative, directly observable 
splitting parameter which 
comprises both the magnon eigenenergy and its magnetic moment and possesses the $g$-wave symmetry in both easy-axial and easy-planar cases.  
%
%
%
%
The associated altermagnetic domain walls in easy-axial CrSb possess a net magnetization
with an amplitude that depends on their orientation.
%

\end{abstract}

\maketitle

\section{Introduction}

The observation of unconventional time reversal symmetry breaking in several collinear compensated magnets
\cite{Smejkal20,Naka19,Yuan20,Naka21,Guo23} has lead to the recent discovery of atlermagnets (AMs) \cite{Smejkal22a,Smejkal22b}, a new third class of collinear compensated magnetism rigorously delimited and classified by spin symmetries. AMs exhibit unconventional spin order with \emph{d, g} or \emph{i}-partial wave parity in the  band structure. 
%
This anisotropic order gives rise to a series of unusual electric and optical properties, 
including the possibility of an anomalous Hall effect governed by the direction of the antiferromagnetic N{\'e}el vector~\cite{Smejkal20,Smejkal22,Sato24}. 
Recently, it has also been established that altermagnetism has very interesting repercussions for other type of  charge neutral magnetic elementary excitations. In $d$-wave AMs, with for instance tetragonal rutile structures, a $d$-wave splitting of the {\it magnon} branches arises, which induces several novel physical effects: a net magnetization induced by nonuniformities of the N{\'e}el vector~\cite{Gomonay24a}, deformation and Walker breakdown for moving domain walls~\cite{Gomonay24a}, fluctuation-induced piezomagnetism~\cite{Consoli21,Yershov24b}, spin currents induced by the temperature gradients~\cite{Yershov24b} and a curvature-induced net magnetization in AM films~\cite{Yershov25}.

The experimentally very fertile hexagonal altermagnetic materials do not exhibit $d$ but $g$-wave symmetry, in particular metallic CrSb~\cite{Smejkal22b,Reimers24,Li24,Yang25,Zeng24a,Ding24a}, semiconducting MnTe~\cite{Smejkal22b,Mazin23,Krempasky24,Lee24,Osumi24}, and semimetallic VNb$_3$S$_6$~\cite{Smejkal22b}. These hexagonal AMs with NiAs-type structure belong to the crystallographic group 6/mmm. Their $g$-wave altermagnetic splitting of the electron bands has been convincingly demonstrated experimentally by angle-resolved photoemission spectroscopy, in both MnTe~\cite{Krempasky24,Lee24,Osumi24} and CrSb~\cite{Reimers24,Li24,Yang24}, which in addition harbours electronic Weyl nodes and topological surface states~\cite{Li24}.

\begin{figure}
	\includegraphics[width=\columnwidth]{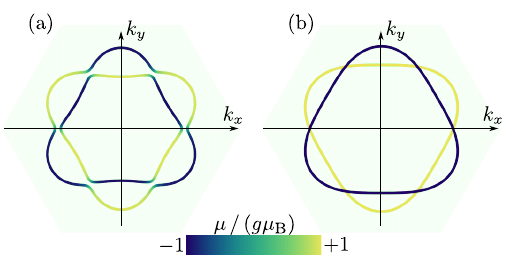}
	\caption{Schematic representation of the isolines of constant energy $\omega_\nu=\text{const}$ for MnTe (panel a) and CrSb (panel b) made for a constant $k_z>0$. Distribution of the amplitude of the magnon magnetic moment $\vec{\mu}$ along the isolines is shown by the color scheme.  In both cases $\vec{\mu}||\vec{n}_0$ where $\vec{n}_0$ is the ground state N{\'e}el vector. For details of the magnon magnetic moment computation, see Section~\ref{sec:mmm}. The green background hexagon indicates the size of the first Brillouin zone.
    }\label{fig:CrSb-vs-MnTe1}
\end{figure}

Here we show that the purely magnetic excitations and spin-textures of these $g$-wave AMs have properties very distinct from $d$-wave ones. Depending on the magnetocrystalline anisotropy being easy-axial (as for CrSb) or easy-planar (as for MnTe) magnon spectra are drastically different, see Fig.~\ref{fig:CrSb-vs-MnTe1}.
While the splitting of the magnon bands of CrSb is not affected by the anisotropy and possesses  a $g$-wave symmetry, as for the nonrelativistic case (see Fig.~\ref{fig:CrSb-vs-MnTe1}b),  the splitting of the magnon bands of MnTe is strongly affected by the anisotropy and, strictly speaking, does not possess $g$-wave symmetry (see Fig.~\ref{fig:CrSb-vs-MnTe1}a)
because of the presence of a finite gap between the branches at any momentum.
 
We will show that each chiral magnon branch in CrSb carries a quantized magnetic moment that is parallel or antiparallel to the ground state N{\'e}el vector $\vec{n}_0$, which points along its easy axis. 
Also, magnons in MnTe carry a magnetic moment $\vec{\mu}$ collinear to the $\vec{n}_0$ oriented in-plane, but the magnitude and orientation of $\vec{\mu}$ depend on momentum, see Fig.~\ref{fig:CrSb-vs-MnTe1}a. The magnetic moment in MnTe is an altermagnetic property and is not expected for simple conventional easy-planar antiferromagnets.  For each magnon branch, the distribution of the magnon magnetic moment within the first Brillouin zone possesses the $g$-wave symmetry. 
%
%
We show that in spite of this difference in the properties of the chiral magnons for CrSb and MnTe, the presence of the magnon moment allows the definition of an universal $g$-wave symmetry characteristic
\begin{equation}\label{eq:lambda}
    \lambda=\frac{1}{\omega_0g\mu_\textsc{b}}\sum\limits_{\nu=\pm}(\mu_\nu)_{\textsc{gs}}\,\omega_\nu
\end{equation}
that comprises both cases and generalizes the concept of spin-splitting of altermagnetic magnon bands beyond the non-relativistic limit. Here $\nu$ numerates the magnon branches, $(\mu_\nu)_{\textsc{gs}}=\vec{\mu}_\nu\cdot\vec{n}_0$ is the magnon magnetic moment along the ground state, $\omega_0$ is some characteristic frequency, $g\approx2$ is the electron $g$-factor, and $\mu_{\textsc{b}}$ is the Bohr magneton.

From analyzing the corresponding AM continuum Hamiltonian we find that magnetic domain wall (DW) textures in CrSb possess a net magnetization, in analogy to $d$-wave altermagnets~\cite{Gomonay24a}, with an amplitude that depends on the DW orientation in the crystal, and we determined twelve orientations that correspond to the maximal magnetization.

\begin{figure}
    \centering
    \includegraphics[width=\columnwidth]{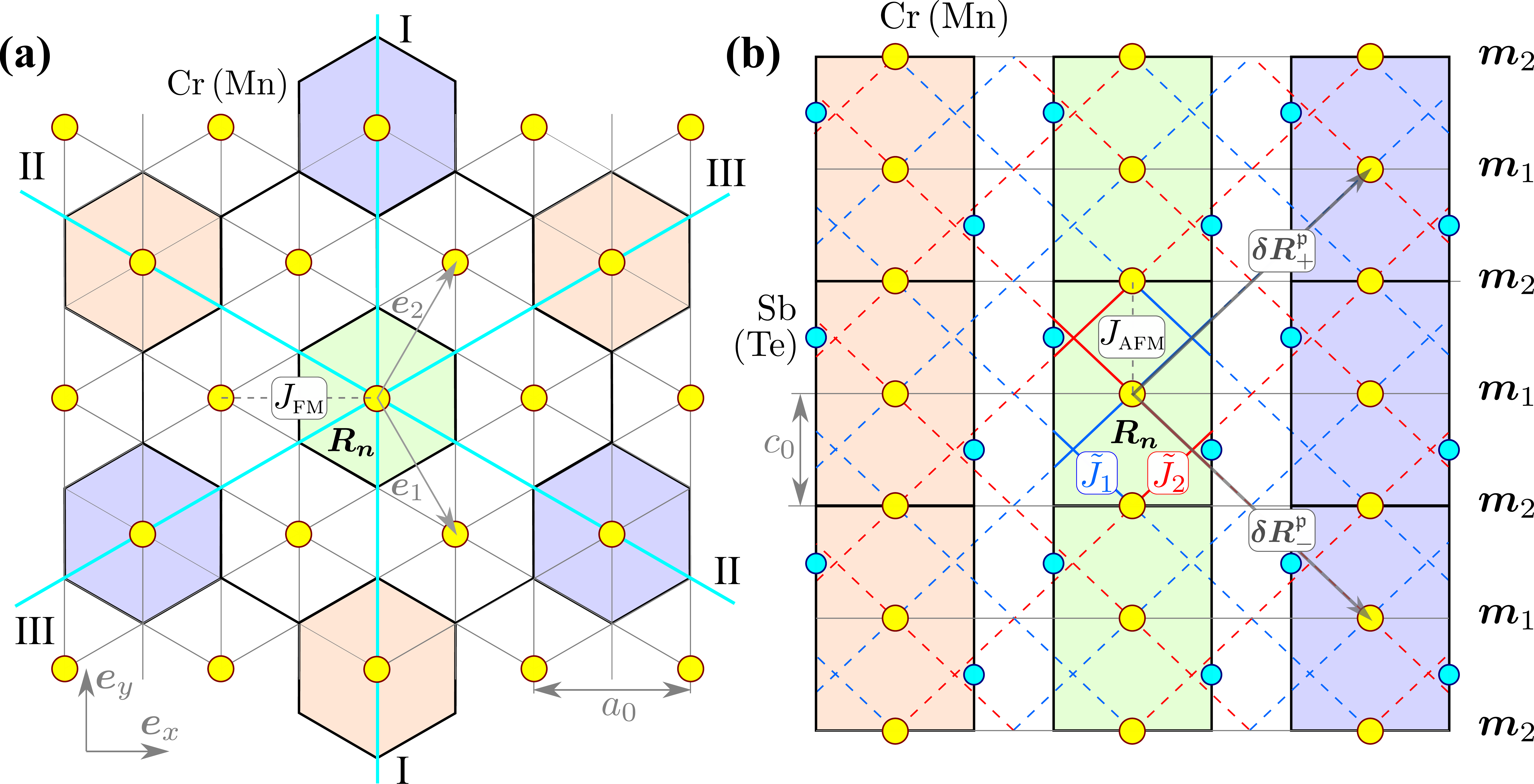}
    \caption{Schematic crystal structure and exchange interactions in hexagonal CrSb and MnTe-type altermagnets.  (a) the (001) plane of magnetic atoms (yellow discs: Cr or Mn) which form a triangular lattice with basis $\vec{e}_{1}$, $\vec{e}_{2}$. Thin lines connect next-nearest neighbors while thick lines border the hexagonal 2D Wigner-Seitz cells. Ferromagnetic Heisenberg exchange of strength $J_{\textsc{fm}}$ couples nearest neighbors in each layer. 
    (b) cross-section of one of the vertical planes I, II or III. The antiferromagnetic exchange  $J_{\textsc{afm}}$ couples nearest magnetic atoms in two neighboring layers at distance $c_0$. Altermagnetic Heisenberg bonds $\tilde{J}_{1(2)}$ are shown by blue (red) lines.  The structure of the altermagnetic bonds is unique for all planes I, II, and III if the orientation of the cross-section (b) is consistent with the coloring of the unit cells (orange-green-magenta).
    }\label{fig:model}
\end{figure}

\section{Magnetic Hamiltonian and its ground states}

The magnetic structure of CrSb and MnTe consists of ferromagnetic (001) planes which are coupled antiferromagnetically, see Fig.~\ref{fig:model}. Their magnetic  Hamiltonian comprises five contributions: $\mathcal{H}=\mathcal{H}_{\textsc{afm}}+\mathcal{H}_{\textsc{fm}}+\mathcal{H}_{\textsc{alt}}+\mathcal{H}_{\textsc{an}}+\mathcal{H}_{\textsc{z}}$. Here $\mathcal{H}_{\textsc{afm}}$ is the antiferromagnetic Heisenberg exchange interaction between the nearest (001) planes illustrated by the bond $J_{\textsc{afm}}$ in Fig.~\ref{fig:model}(b). $\mathcal{H}_{\textsc{fm}}$ takes into account the nearest-neighbor ferromagnetic exchange within each (001) plane, see the bond $J_{\textsc{fm}}$ in Fig.~\ref{fig:model}(a). The altermagnetic properties are encoded in $\mathcal{H}_{\textsc{alt}}$ employing the  additional Heisenberg exchanges of strengths $\tilde{J}_{1,2}$ shown in Fig.~\ref{fig:model}(b) by red and blue lines. Orientations of the altermagnetic bonds are determined by the displacement vectors $\vec{\delta R}_{\pm}^{\mathfrak{p}}$ with $\mathfrak{p}\in\{\textsc{i},\textsc{ii},\textsc{iii}\}$ denoting the cross-section plane comprising the displacement vector, see cyan lines in Fig.~\ref{fig:model}(a). Here $\vec{\delta R}_{\pm}^{\text{I}}=-\vec{e}_1+\vec{e}_2\pm\vec{e}_3$,  $\vec{\delta R}_{\pm}^{\text{II}}=2\vec{e}_1+\vec{e}_2\pm\vec{e}_3$,  $\vec{\delta R}_{\pm}^{\text{III}}=-\vec{e}_1-2\vec{e}_2\pm\vec{e}_3$ with $\vec{e}_1=\frac{a_0}{2}\left(\vec{e}_x-\sqrt{3}\vec{e}_y\right)$, $\vec{e}_2=\frac{a_0}{2}\left(\vec{e}_x+\sqrt{3}\vec{e}_y\right)$, and $\vec{e}_3=2c_0\vec{e}_z$ being the basis vectors. Note the very different relative locations of the red and blue bonds concerning the nonmagnetic atoms Sb or Te indicated by blue dots: red bonds pass in close proximity to the nonmagnetic atoms. Due to the superexchange interaction realized through the nonmagnetic atoms we have $\tilde{J}_{1}\ne\tilde{J}_{2}$~\cite{Liu24b}. In the following we introduce the symmetrization $\tilde{J}_1=\tilde{J}-\delta\tilde{J}$, $\tilde{J}_2=\tilde{J}+\delta\tilde{J}$ and show that the altermagnetic effects are proportional to $\delta\tilde{J}$. The explicit expressions for the exchange contributions $\mathcal{H}_{\textsc{afm}}$, $\mathcal{H}_{\textsc{fm}}$, and $\mathcal{H}_{\textsc{alt}}$ are presented in Eqs.~\eqref{eq:Hafm}, \eqref{eq:Hfm}, and \eqref{eq:Halt}, respectively. Additionally, we take into account the uniaxial anisotropy $	\mathcal{H}_{\textsc{an}}=-K\sum_{\vec{R}_{\vec{n}}}\left[m_{1z}^2(\vec{R}_{\vec{n}})+m_{2z}^2(\vec{R}_{\vec{n}}+c_0\vec{e}_z)\right]$ with $K>0$ and $K<0$ for CrSb and MnTe, respectively. Here $\vec{m}_\nu(\vec{r}_{\vec{n}})$ with $\nu=1,2$ is a unit vector showing the direction of the magnetic moment of $\nu$-th sublattice located in position $\vec{r}_{\vec{n}}$. The lattice vectors $\vec{R}_{\vec{n}}=n_1\vec{e}_1+n_2\vec{e}_2+n_3\vec{e}_3$ with $n_i\in\mathbb{Z}$ numerate the hexagonal unit cells the primitive magnetic unit cells in the form of hexagonal prisms bordered by thick lines in Fig.~\ref{fig:model}.
The interaction with external magnetic field $\vec{B}=B\vec{e}_z$ is captured by the Hamiltonian $\mathcal{H}_{\textsc{z}}=-B\mu_s\sum_{\vec{R}_{\vec{n}}}\left[m_{1z}(\vec{R}_{\vec{n}})+m_{2z}(\vec{R}_{\vec{n}}+c_0\vec{e}_z)\right]$, where $\mu_s$ is magnetic moment of a magnetic atom. 

Parameters of the introduced model are listed in Table~\ref{tab:params-discr}.
\begin{table}[h]
    \centering
    \begin{tabular}{|c||c|c|}
        \hline
         &  CrSb & MnTe\\
        \hline\hline
        $J_{\textsc{afm}}$ & 35.86 meV  &  3.99 meV  \\
        \hline
        $J_{\textsc{fm}}$ & 13.53 meV  &  0.12 meV  \\
        \hline
        $\tilde{J}$ & $0.5$ meV  &  0.023 meV  \\
        \hline  
        $\delta\tilde{J}$ & $0.5$ meV  &  0.045 meV  \\
        \hline
        $K$ & 0.49 meV
        & $-0.048$ meV~\cite{Liu24b}\\
        \hline
        $\mu_s$ & $3 \mu_\textsc{b}$ \cite{Polesya11} &  $5\mu_\textsc{b}$ \cite{Mazin24a} \\
        \hline  
        $a_0$ &  4.12 \AA~\cite{Takei63}    &  4.17 \AA~\cite{Greenwald53,Villars23}  \\
        $c_0$ & 2.73 \AA~\cite{Takei63} & 3.375 \AA~\cite{Greenwald53,Villars23} \\
        \hline
    \end{tabular}
    \caption{Parameters of the discrete model estimated for CrSb and MnTe. The values for the exchange and anisotropy constants of CrSb and MnTe are taken from own 
    DFT calculations \footnote{From DFT calculation we obtain $\tilde{J}_1+\tilde{J}_2\approx1$ meV, and assume that $\tilde{J}_1=0$.} and from Ref.~\onlinecite{Liu24b}, respectively. }
    \label{tab:params-discr}
\end{table}
This model is minimal in the sense that further (longer-range) magnetic coupling might be present. However, these will not change the AM symmetries nor the mathematical structure of low-energy continuum description that we will derive next, in order to investigate magnetic domain wall properties.
Since the antiferromagnetic exchange strongly dominates the rest of the interactions for both CrSb and MnTe, the ground state is a uniform AFM collinear magnetization oriented parallel and perpendicular to the anisotropy axis [001] for CrSb and MnTe, respectively. An applied magnetic field $\vec{B}=B\vec{e}_z$ does not change the ground state of CrSb $\vec{m}_\nu^0=(-1)^{\nu-1}\vec{e}_z$, if  $|B|<B_{sf}$ where 
$B_{sf}\approx\sqrt{8KJ_{\textsc{afm}}}/\mu_s\approx68$~T is the spin-flop field~\footnote{Because of a large estimated value of $B_{sf}$, we do not consider the spin-flop phase of CrSb in this paper.}. Since in our case $K\ll J_{\textsc{afm}}$, the field region in the vicinity of $B_{sf}$ which corresponds to the  metastable collinear state~\cite{Bogdanov07} is negligibly narrow. 

In the case of MnTe, the applied field leads to the canting of the magnetic moments resulting in the net magnetization along the field. Thus, the ground state for MnTe is $\vec{m}_\nu^0=(-1)^{\nu-1}\sin\theta_0(\vec{e}_x\cos\phi_0+\vec{e}_y\sin\phi_0)+\cos\theta_0\vec{e}_z$, where $\phi_0$ is an arbitrary constant and $\cos\theta_0=B/B_0$ with 
$B_0\approx4J_{\textsc{afm}}/\mu_s$ being the field of spin-flip. Here we neglect possible weak in-plane anisotropies which may fix the orientation of $\vec{m}_\nu^0$ in a real crystal. 


\begin{figure*}
	\includegraphics[width=\textwidth]{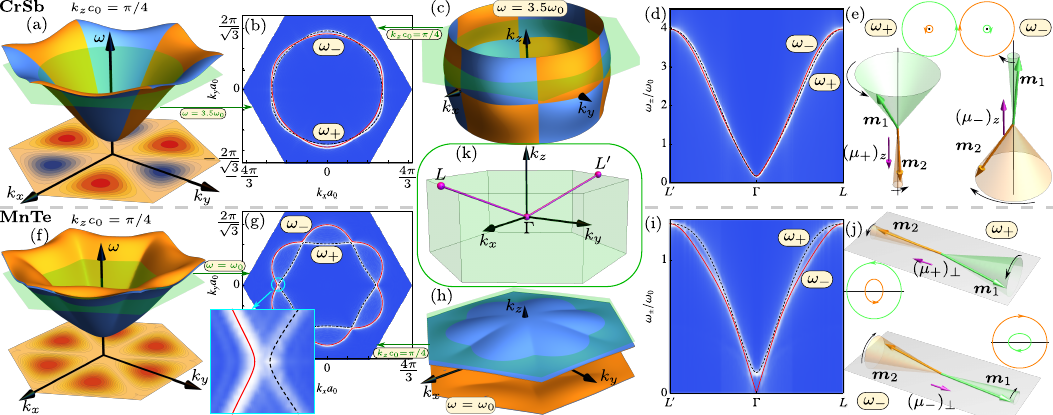}
	\caption{ Comparison of the magnon spectra \eqref{eq:disp-CSb} for CrSb (top row) and \eqref{eq:disp-MnTe} for MnTe (bottom row). The figures are built for the values of the parameters listed in Table~\ref{tab:params-discr} and for $B=0$. Panels (a) and (f) show the evolution of two branches $\omega_{+}$ (orange) and $\omega_{-}$ (blue) within the 1st Brillouin zone (1.BZ) for $k_z=\text{const}$. The value of the splitting $\Delta\omega=\omega_{+}-\omega_{-}$ is shown by the color coding at the bottom. Surfaces of constant energy $\omega=\mathrm{const}$ are shown on panels (c) and (h) for CrSb and MnTe, respectively. Panels (b) and (g) aggregate panels (a,c) and (f,h), respectively, showing the surfaces of the constant energy for the constant $k_z$. On panels (b,d,g,i), the analytical dispersion (lines) is compared to the spectra extracted from the spin-lattice simulations (color density). Panel (k) shows the 1.BZ and the path $L'\Gamma L$ along which we computed the dispersions presented on panels (d,i). Panels (e) and (j) show precession of the neighboring magnetic moments of different sublattices  for a spin wave with wave-vector $(k_xa_0, k_ya0, k_zc_0)=(0,\pi/\sqrt{3},\pi/4)$ for the case of CrSb and MnTe, respectively. The insets show the trajectories swept by the ends of the vectors $\vec{m}_\nu$.}\label{fig:dispersion-CrSb-MnTe}
\end{figure*}

\section{Dispersion and polarization of spin-wave eigen-modes}

Dynamics of each magnetic moment $\vec{m}_\nu(\vec{R}_{\vec{n}})$  is described by the Landau-Lifshitz equation 
\begin{equation}\label{eq:LL}
	\dot{\vec{m}}_{\nu}(\vec{R}_{\vec{n}})=\frac{\gamma}{\mu_s}\left[\vec{m}_{\nu}(\vec{R}_{\vec{n}})\times\frac{\partial\mathcal{H}}{\partial\vec{m}_{\nu}(\vec{R}_{\vec{n}})}\right],
\end{equation}
where $\gamma=g\mu_{\textsc{b}}/\hbar>0$ is the electron gyromagnetic ratio. The set of equations \eqref{eq:LL} coupled via Hamiltonian $\mathcal{H}$ describes the collective dynamics of all magnetic moments. 
To describe the dynamics of deviations from a ground state $\vec{m}^0_\nu=\sin\Theta_\nu(\vec{e}_x\cos\Phi_\nu+\vec{e}_y\sin\Phi_\nu)+\cos\Theta_\nu\vec{e}_z$, it is convenient to introduce the complex-valued function $\psi_\nu$~\cite{Tyablikov75}:
\begin{equation}\label{eq:Tyabl}
 \vec{m}_\nu=\vec{m}_\nu^0\left(1-|\psi_\nu|^2\right)+\sqrt{2-|\psi_\nu|^2}(\vec{T}_\nu\psi_\nu+\vec{T}^*_\nu\psi^*_\nu),
\end{equation}
where the form of null vector $\vec{T}_\nu=\frac12(\cos\Theta_\nu\cos\Phi_\nu-i\sin\Phi_\nu)\vec{e}_x+\frac12(\cos\Theta_\nu\sin\Phi_\nu+i\cos\Phi_\nu)\vec{e}_y-\frac12\sin\Theta_\nu\vec{e}_z$ guarantees fulfillment of the constraint $|\vec{m}_\nu|=1$~\footnote{One should take into account the following properties $\vec{m}_\nu^0\cdot\vec{T}_\nu=0$, $\vec{T}_\nu\cdot\vec{T}_\nu=0$, and $\vec{T}_\nu\cdot\vec{T}^*_\nu=1/2$.}. Note that for the particular case of the CrSb ground state, Eq.~\eqref{eq:Tyabl} becomes a classical analog of the Holstein-Primakoff representation~\cite{Holstein40}. In terms of $\psi_\nu$-functions, the Landau-Lifshitz equations~\eqref{eq:LL} take the Schr{\"o}dinger-like form
\begin{equation}\label{eq:LL-psi}
i\dot{\psi}_\nu(\vec{R}_{\vec{n}})=\frac{\gamma}{\mu_s}\frac{\partial\mathcal{H}}{\partial\psi^*_\nu(\vec{R}_{\vec{n}})}.
\end{equation}


\subsection{Magnon dispersion relation}
The linearization of Eqs.~\eqref{eq:LL-psi} with respect to small $\psi_\nu$ and the subsequent solution of the eigenvalue problem for a periodic lattice enables us to obtain the dispersion relation for CrSb (see Appendix~\ref{app:magnon-spectrum})
\begin{equation}\label{eq:disp-CSb}
\omega_{\pm}^{\text{CrSb}}\!=\!\omega_0\!\left[\sqrt{(F_{\vec{k}}+\kappa)^2-\cos^2(k_zc_0)}\pm\delta\varepsilon\Omega_{\vec{k}}^{-}\right]\!\pm\gamma B,
\end{equation}
where $F_{\vec{k}}=1+\eta\Omega_{\vec{k}}^{\textsc{fm}}-\varepsilon\Omega_{\vec{k}}^{+}$. The dispersions $\Omega_{\vec{k}}^{\textsc{fm}}=\sin^2(\vec{k}\cdot\vec{e}_1/2)+\sin^2(\vec{k}\cdot\vec{e}_2/2)+\sin^2(\vec{k}\cdot(\vec{e}_1+\vec{e}_2)/2)$, and $\Omega_{\vec{k}}^{\pm}=\sum_{\mathfrak{p}\in\{\textsc{i,ii,iii}\}}\left[\sin^2(\vec{k}\cdot\vec{\delta R}_+^{\mathfrak{p}}/2)\pm\sin^2(\vec{k}\cdot\vec{\delta R}_-^{\mathfrak{p}}/2)\right]$ originate from the ferromagnetic and altermagnetic contributions, respectively. The parameters are defined as: $\omega_0=2\gamma J_{\textsc{afm}}/\mu_s$
, $\kappa=K/J_{\textsc{afm}}$
, $\eta=4J_{\textsc{fm}}/J_{\textsc{afm}}$
, $\varepsilon=2\tilde{J}/J_{\textsc{afm}}$
, and $\delta\varepsilon=2\,\delta\tilde{J}/J_{\textsc{afm}}$
. 
In the vicinity of $\Gamma$-point, 
dispersion relation \eqref{eq:disp-CSb} simplifies to 
\begin{equation}\label{eq:disp-Gamma}
	\begin{split}
\omega^{\text{CrSb}}_{\pm}&\approx\sqrt{\omega_{\textsc{afmr}}^2+\mathfrak{c}_\perp^2k_\perp^2+\mathfrak{c}_z^2k_z^2+\mathcal{O}(k^4)}\\
 &\pm\left[\Lambda k_zk_y(k_y^2-3k_x^2)+\gamma B\right].
 \end{split}
\end{equation}
Here $k_\perp^2=k_x^2+k_y^2$ and $\omega_{\textsc{afmr}}=\gamma B_{sf}\approx1.2\times10^{13}$~rad$/$s (1.9~THz) is the frequency of the uniform antiferromagnetic resonance. $\mathfrak{c}_\perp=\omega_{\textsc{afmr}}\ell_\perp$ and $\mathfrak{c}_z=\omega_{\textsc{afmr}}\ell_z$ are the magnon speeds in $xy$-plane and along $z$-axis, respectively, where the typical length scales are $\ell_\perp=a_0[(3J_{\textsc{fm}}-9\tilde{J})/(2K)]^{1/2}\approx2.5$~nm and $\ell_z=c_0[(J_{\textsc{afm}}-24\tilde{J})/(2K)]^{1/2}\approx1.3$~nm. Parameter $\Lambda=3\sqrt{3}\frac{\gamma}{\mu_s}\delta\tilde{J}\,a_0^3c_0$ represents the altermagnetic strength and determines the splitting between branches. The latter is of 4th order in $k$, in contrast to the 2nd order splitting in $d$-wave altermagnets~\cite{Gomonay24a}.
\begin{figure}
   \centering
   \includegraphics[width=0.95\columnwidth]{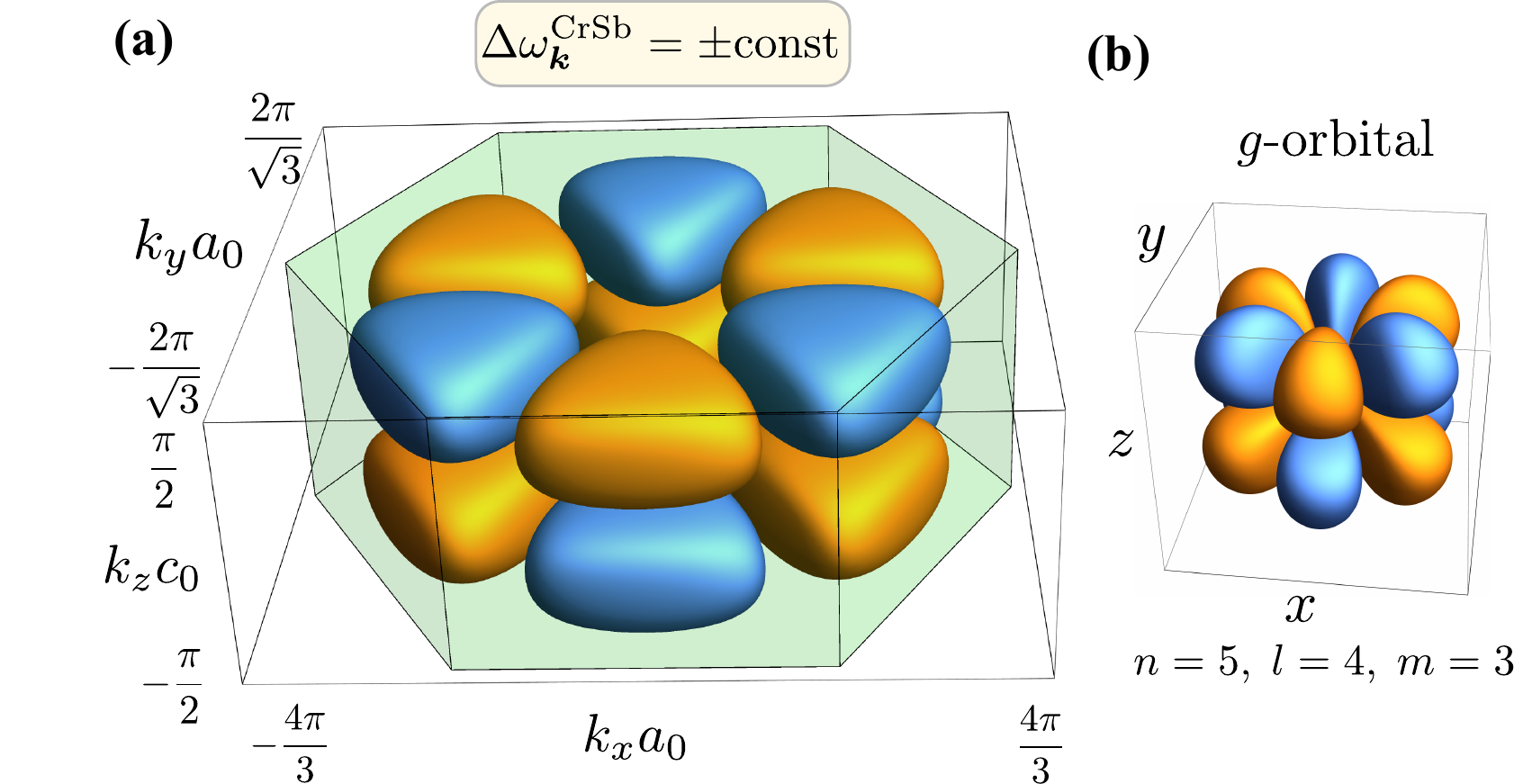}
   \caption{(a) Surfaces of constant magnon splitting $\Delta\omega^{\text{CrSb}}_{\vec{k}}=\pm8\gamma\delta\tilde{J}/\mu_s$ of the magnon branches of CrSb in the first BZ compared to (b) the hydrogen $g$-orbital with the corresponding quantum numbers $(n,l,m)$. Orange and blue color correspond to $\Delta\omega^{\text{CrSb}}_{\vec{k}}>0$ and $\Delta\omega^{\text{CrSb}}_{\vec{k}}<0$, respectively.}
   \label{fig:splitting-CrSb}
\end{figure}

The magnon spectra in CrSb have the same properties as the electron spectra in this material  
in terms of spin-splitting of the dispersive branches~\cite{Reimers24,Li24,Yang25,Zeng24a,Ding24a}. For $B=0$, the altermagnetism, which manifests through the parameter $\delta\tilde{J}$, lifts the Kramers' degeneracy of the magnon branches, see the upper row in Fig.~\ref{fig:dispersion-CrSb-MnTe}. Note, that the surfaces of constant energies for magnons shown in Fig.~\ref{fig:dispersion-CrSb-MnTe}(c) are similar to the Fermi surfaces recently computed for electrons in CrSb~\cite{Li24,Ding24a}. The magnon branches of CrSb possess splitting of a bulk $g$-wave symmetry: $\Delta\omega^{\text{CrSb}}_{\vec{k}}=2\omega_0\,\delta\varepsilon\,\Omega_{\vec{k}}^-=-\frac{16\gamma\,\delta\tilde{J}}{\mu_s}\times\sin(2k_zc_0)\sin\left(\frac{\sqrt{3}}{2}k_ya_0\right)\bigl[\cos\left(\frac{3}{2}k_xa_0\right)-\cos\left(\frac{\sqrt{3}}{2}k_ya_0\right)\bigr]$,
see Fig.~\ref{fig:splitting-CrSb} which demonstrates the similarity of the surfaces of constant splitting $\Delta\omega^{\text{CrSb}}_{\vec{k}}=\pm\text{const}$ with a $g$-orbital of a hydrogen atom. The splitting is absent if the $\vec{k}$-vector lies in either the horizontal plane (001) or in one of three vertical nodal planes containing vector $\vec{e}_1$, or $\vec{e}_2$ or $\vec{e}_1+\vec{e}_2$, see Fig.~\ref{fig:model}. The maximal splitting ($6.9$~meV) is achieved when $\vec{k}$-vector lies in one of the vertical planes I, II or III (see Fig.~\ref{fig:model}) making angle $\vartheta_{\text{max}}=\arctan\frac{16c_0}{3\sqrt{3}a_0}\approx63.9^\circ$ with the vertical direction [001] and having the absolute value $k_{\text{max}}=\pi[\frac{16}{27a_0^2}+\frac{1}{16c_0^2}]^{1/2}\approx0.65$~rad/\AA. Note that the value of $\Delta\omega^{\text{CrSb}}_{\vec{k}}$ is determined by the altermagnetic strength $\delta\tilde{J}$ and is not influenced by the anisotropy.

Similarly, considering the corresponding ground state, we obtain the dispersion relation for the magnons in MnTe (see Appendix~\ref{app:magnon-spectrum})
%
%
\begin{align}\label{eq:disp-MnTe}
    \nonumber&\frac{\omega^{\text{MnTe}}_{\pm}}{\omega_0}\!=\!\Biggl\{\!F_{\vec{k}}(F_{\vec{k}}\!+\!\kappa\sin^2\theta_0)\!+\!\cos2\theta_0\cos^2(k_zc_0)\!+\!(\delta\varepsilon\Omega^-_{\vec{k}})^2\\
    \nonumber&\pm\!\Bigl[\cos^2(k_zc_0)\left\{\left[\kappa\!+\!\cos^2\theta_0(2F_{\vec{k}}\!-\!\kappa)\right]^2\!-\!4(\delta\varepsilon\Omega^-_{\vec{k}})^2\sin^2\theta_0\right\}\\
    &+(\delta\varepsilon\Omega^-_{\vec{k}})^2(2F_{\vec{k}}+\kappa\sin^2\theta_0)^2\Bigr]^{\frac12}\Biggr\}^{\frac12},
\end{align}
where $\kappa=|K|/J_{\textsc{afm}}$ and definitions of the constants $\omega_0$, $\eta$, $\varepsilon$, $\delta\varepsilon$ are the same as in~\eqref{eq:disp-CSb}. Due to the different ground states, the spectra of CrSb and MnTe are drastically different. The ground state of MnTe is continuously degenerate with respect to its orientation within the easy plane. The latter circumstance leads to the emergence of the Goldstone mode, see the spectra behavior in $\Gamma$-point in Fig.~\ref{fig:dispersion-CrSb-MnTe}i. In contrast to CrSb, the splitting $\Delta\omega^{\text{MnTe}}_{\vec{k}}$ of the magnon branches of MnTe is strongly influenced by the anisotropy and, strictly speaking, does not possess $g$-wave symmetry. The magnon branches $\omega^{\text{MnTe}}_{\pm}$ are not degenerate even if $B=0$ and $\delta\tilde{J}=0$. The introduction of the altermagnetism ($\delta\tilde{J}\ne0$) leads to the 6-fold modulations of the magnon branches, see the bottom row in Fig.~\ref{fig:dispersion-CrSb-MnTe}. 
Nevertheless, the magnon branches do not intersect and the splitting $\Delta\omega^{\text{MnTe}}_{\vec{k}}$ does not alternate sign. However, for the particular values of the parameters listed in Table~\ref{tab:params-discr} the distance between two magnon branches can be rather small (see Fig.~\ref{fig:dispersion-CrSb-MnTe}g) and, therefore, may be difficult to resolve experimentally~\cite{Liu24b}. 

In the limit case $K=0$, $B=0$, the dispersion \eqref{eq:disp-MnTe} is simplified to $\omega^{\text{MnTe}}_{\pm}=\omega_0[\sqrt{F_{\vec{k}}^2-\cos^2(k_zc_0)}\pm|\delta\varepsilon\Omega_{\vec{k}}^-|]$. In the considered limit, $\omega^{\text{MnTe}}_{\pm}$ reproduces $\omega^{\text{CrSb}}_{\pm}$ in \eqref{eq:disp-CSb}, but $\Delta\omega^{\text{MnTe}}_{\vec{k}}=|\Delta\omega^{\text{CrSb}}_{\vec{k}}|$. However, both spectra become identical if the magnetic moment of the magnon is used as a criterion for distinguishing modes, see Fig.~\ref{fig:dispersion-CrSb-MnTe_vs_kappa}d,h. 
The evolution of both spectra with $K$ can be traced in Fig.~\ref{fig:dispersion-CrSb-MnTe_vs_kappa}.

Solving the eigenvalue problem, we reconstruct the dynamics of the magnetic moments $\vec{m}_\nu(\vec{R}_{\vec{n}})$ using eigenvectors and Eq.~\eqref{eq:dm-sw}. The obtained dynamics of two neighboring moments $\vec{m}_1$ and $\vec{m}_2$ is shown in Fig.~\ref{fig:dispersion-CrSb-MnTe}e. The dynamics is typical for easy-axial antiferromagnets~\cite{Rezende19}, i.e. moments $\vec{m}_{\nu}$ demonstrate circular precession about the anisotropy easy-axis shown by the vertical gray line. For a given spin wave, the direction of precession is the same for both sublattices $\vec{m}_1$ and $\vec{m}_2$, however, it is opposite for the spin waves belonging to different branches $\omega_+$ and $\omega_-$. The latter allows us to characterize magnons by \emph{chirality}, that is, by the direction of precession clockwise (branch $\omega_+$) or counterclockwise (branch $\omega_-$) relative to the direction $\vec{m}_1^0$. In the branch $\omega_+$ ($\omega_-$), the magnetic moment $\vec{m}_2$ ($\vec{m}_1$) precesses in the direction opposite to the natural precession direction, which would appear in the absence of the AFM coupling between the sublattices. Thus, one can say that the moment $\vec{m}_1$ enforces the unfavorable direction of precession for the moment $\vec{m}_2$ for the branch $\omega_+$ and vice-versa for the branch $\omega_-$. The amplitudes of the unfavorable precessions are smaller and the difference in the precession amplitudes of moments $\vec{m}_1$ and $\vec{m}_2$ gives rise to the uncompensated magnetic moment $(\mu_\pm)_z$. The latter has opposite directions for different branches and, therefore, it correlates with magnon chirality, i.e. the magnons of opposite chiralities have opposite magnetic moments. In the next section we provide a rigorous calculation of the magnon magnetic moment. 

\begin{figure*}
    \centering
    \includegraphics[width=\textwidth]{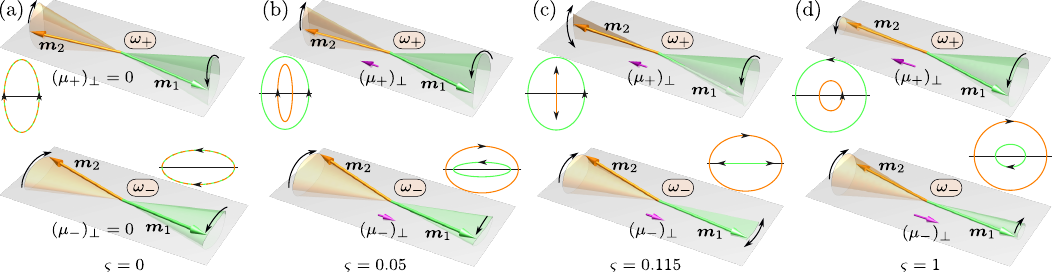}
    \caption{The altermagnetically induced emergence of the chirality and magnetic moment in spin-waves for MnTe. The precession dynamics is reconstructed by means of Eq.~\ref{eq:dm-sw} for the wave-vector $(k_xa_0, k_ya_0, k_zc_0)=(0,\pi/\sqrt{3},\pi/4)$. To illustrate the role of the altermagnetism, we introduce the scaling factor $\varsigma$ for the altermagnetic parameters $\varepsilon$ and $\delta\varepsilon$ of MnTe and show the precession dynamics for $(\varepsilon',\delta\varepsilon')=\varsigma(\varepsilon,\delta\varepsilon)$ with $\varsigma$ varying from 0 to 1. The easy-plane is shown by gray.}
    \label{fig:precess-MnTe}
\end{figure*}

The altermagnetism of CrSb does not change qualitatively the precession dynamics described above. This is in strong contrast to MnTe, where aletrmagnetism drastically influences the precession of the magnetic moments. Similarly to all easy-planar AFMs, magnetic moments of MnTe demonstrate elliptical precession with the semi-major axes parallel and perpendicular to the easy-plane for the lower ($\omega_-$) and higher ($\omega_+$) branches, respectively, see Fig.~\ref{fig:dispersion-CrSb-MnTe}j. For vanishing altermagnetic interactions, see Fig.~\ref{fig:precess-MnTe}a, the precession directions of $\vec{m}_1$ and $\vec{m}_2$ are opposite, meaning that this type of dynamics can not be characterized by chirality. Each magnetic moment precesses in its favourite direction, and the precession amplitudes of $\vec{m}_1$ and $\vec{m}_2$ are equal~\cite{Rezende19}. The latter results in zero averaged magnetic moment. Thus, magnons in easy-planar AFMs, are not chiral and do not possess a magnetic moment. With the increase of the altermagnetic parameter, one of the moments changes the direction of the precession by passing through the linear polarization regime, see Fig.~\ref{fig:precess-MnTe}. Finally, both moments precess in the same direction, which can be associated with a certain chirality. This is the mechanism of the altermagnetically induced emergence of the magnon chirality. Similarly to the case of CrSb, the amplitude of the precession in the unfavorable direction is smaller, this results in the uncompensated averaged magnetic moment $(\mu_\pm)_\perp$. In the next section we rigorously introduce the magnetic moment of magnons and their computation.

\subsection{Magnetic moment of spin-wave eigen-modes}\label{sec:mmm}

\begin{figure}
	\includegraphics[width=\columnwidth]{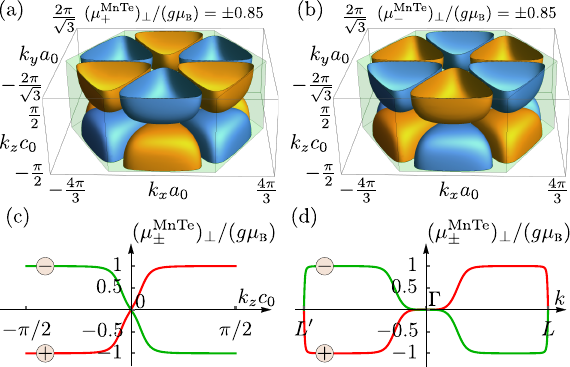}
	\caption{Magnetic moment carried by one magnon in MnTe determined by Eq.~\eqref{eq:mu-exact}. The moment is oriented along the MnTe ground state  and has the amplitude $(\mu^{\text{MnTe}}_\pm)_\perp$, where the subscripts `$+$' or `$-$' correspond to the higher and lower magnon branches, respectively.  Panels (a) and (b) show the distribution of magnon magnetic moment within the BZ by means of the orange (corresponds to $(\mu^{\text{MnTe}}_\pm)_\perp>0$) and blue (corresponds to $(\mu^{\text{MnTe}}_\pm)_\perp<0$) isosurfaces. Panels (a) and (b) correspond to the higher and lower branches, respectively. Panels (c) and (d) show the evolution of the magnetic moment along paths $-\pi/2\le k_zc_0\le\pi/2$ with $k_x a_0=0,\, k_y a_0 = \pi/\sqrt{3}$, and $L'\Gamma L$, respectively. Positive and negative value of $(\mu^{\text{MnTe}}_\pm)_\perp$ corresponds to the magnetic moment orientation along $\vec{m}_1^0$ and $\vec{m}_2^0$, respectively. }\label{fig:magnetization-MnTe}
\end{figure} 

Let us now compute the magnetic moment carried by one magnon in a vanishing magnetic field~\cite{Ashcroft76}
\begin{equation}\label{eq:mu}
    \vec{\mu}_\nu=-\hbar(\partial\omega_\nu/\partial{\vec{B}})|_{\vec{B}=\vec{0}}.
\end{equation}
In this regard, we note the principially different dependence of the spectra of CrSb and MnTe on the magnetic field applied in $z$-direction: while $\omega_{\pm}^{\text{CrSb}}$ is linear in $B$, for MnTe, one has $\omega^{\text{MnTe}}_{\pm}=\omega^{\text{MnTe}}_{\pm}(B^2)$. This difference originates from the different orientation of the magnetic field with respect to the ground state. Consiquently, according to~\eqref{eq:mu}, $(\mu^{\text{CrSb}}_\pm)_z=\mp g\mu_{\textsc{b}}$ while $(\mu^{\text{MnTe}}_\pm)_z=0$. The behavior of the perpendicular components is opposite: while $(\mu^{\text{CrSb}}_\pm)_\perp=0$, magnons in MnTe possess finite magnetic moment oriented along the ground state~\footnote{In order to compute $(\mu_\nu)_{\perp}$ we obtain the dispersion relation for the magnetic field applied within $xy$-plane and use Eq.~\eqref{eq:mu}. For the case of MnTe, we fix the orientation of the ground state by introducing the additional  easy-axial anisotropy $\tilde{K}$ with the easy axis lying within the easy plane. Then we compute magnetic moment \eqref{eq:mu} and take the limit $\tilde{K}\to 0$. For details see Appendix~\ref{app:magnon-spectrum}.}. In the limit $\delta\varepsilon\ll\kappa\ll1$ (equivalnet to $\delta\tilde{J}\ll|K|\ll J_{\textsc{afm}}$)~\footnote{Although this limit is not fulfilled for MnTe, the simplified expression for magnetic moment \eqref{eq:mu-MnTe} is clear and easy to analyze. } we estimate
\begin{equation}\label{eq:mu-MnTe}
    (\mu^{\text{MnTe}}_\pm)_\perp\approx\mp g\mu_{\textsc{b}}\frac{2\delta\varepsilon}{\kappa}\frac{\Omega_{\vec{k}}^-}{\cos(c_0k_z)}\sqrt{F_{\vec{k}}^2-\cos^2(c_0k_z)}.
\end{equation}
For the exact expression, see Eq.~\eqref{eq:mu-exact}. The subscripts `$+$' and `$-$' correspond to the higher and lower magnon branches, respectively. Note that within the accepted assumptions (zero DMI and zero higher-order anisotropies) the magnetic moments of the eigen-modes \eqref{eq:mu-MnTe}, as well as \eqref{eq:mu-exact}, disappear
in the limit $\delta\varepsilon \rightarrow 0$.
The behaviour of the magnon magnetic moments within the BZ is analyzed in Fig.~\ref{fig:magnetization-MnTe}. One can see that for each magnon branch the distribution of the magnon magnetic moment over the BZ has the form of a domain structure with 12 domains:  6 domains are magnetized in direction $\vec{m}_1^0$ (orange isosurfaces), and the other six -- in the opposite direction $\vec{m}_2^0$ (blue isosurfaces), see Fig.~\ref{fig:magnetization-MnTe}a,b. This distribution has a symmetry of a $g$-orbital, compare to Fig.~\ref{fig:splitting-CrSb}b. The transition region between two domains can be treated as a domain wall in $k$-space, it has a kink-like structure, see Fig.~\ref{fig:magnetization-MnTe}c. The magnon magnetic moment vanishes on the side surface of the BZ but not on the top and bottom surfaces, see Fig.~\ref{fig:magnetization-MnTe}c,d. The distribution of $(\mu^{\text{MnTe}}_\pm)_\perp$ qualitatively reproduces distribution of the neutron chiral factor for the magnons in MnTe~\cite{Liu24b}.

It was recently shown that the similar $k$-dependent spin polarization of electronic bands in MnTe occurs due to the spin-orbit coupling~\cite{Krempasky24}. In contrast to the magnetic moment of a magnon, the electron spin has a fixed length, so it necessarily rotates when moving from one domain to another. The latter leads to the appearance of $z$-component of the electron spin polarization~\cite{Krempasky24}. The magnon magnetic moment $\vec{\mu}_\pm^{\text{MnTe}}$, however, is always collinear to the ground state and vanishes on the domain wall.

\begin{figure*}
	\includegraphics[width=0.9\textwidth]{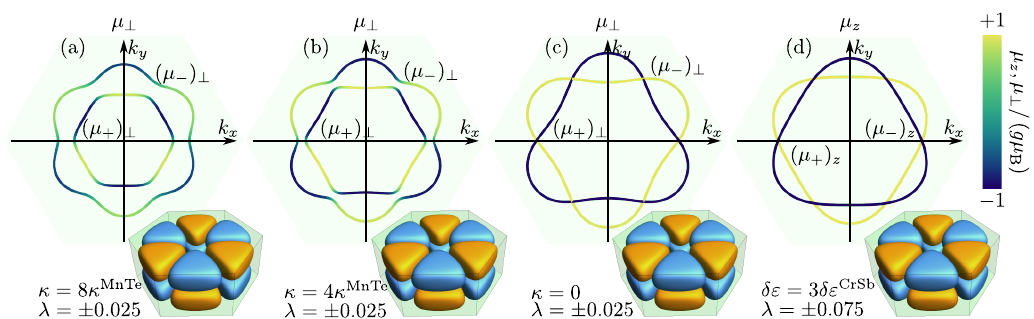}
	\caption{ Distribution of amplitude of the magnon magnetic moment along the isolines of constant energy $\omega_\pm=\text{const}$ for easy-planar (panels a,b,c) and easy-axial (panel d) anisotropies, respectively. All isolines are built for constant $k_z=\pi/(4c_0)$. Panels a-c show evolution of the isolines as anisotropy decreases to zero. Parameters of panels (a)-(c) are the same as for MnTe (see Fig.~\ref{fig:dispersion-CrSb-MnTe}g) but anisotropy. Parameters for panel (d) are the same as for CrSb (see Fig.~\ref{fig:dispersion-CrSb-MnTe}b) but the enhanced altermagnetic strength $\delta\varepsilon\to3\delta\varepsilon$ (to visually increase the altermagnetic splitting). The bottom row demonstrates distribution of the altermagnetic parameter~\eqref{eq:lambda} over the Brillouin zone by means of the isosurfaces $\lambda=\pm\text{const}$. }\label{fig:CrSb-vs-MnTe}
\end{figure*} 

It is instructive to compare the distributions of the magnon magnetic moment along the isolines of the constant energy $\omega_\pm(k_x,k_y,k_z=\text{const})=\text{const}$ for the easy-planar (MnTe) and easy-axial (CrSb) cases, see Figs.~\ref{fig:CrSb-vs-MnTe1},\ref{fig:CrSb-vs-MnTe}. One can see that this distribution is very similar for both cases, except for the fact that the degeneracy points for the easy-axial CrSb are gapped out for the easy-planar MnTe. The gaps, however, vanish for the vanishing easy-planar anisotropy, see Fig.~\ref{fig:CrSb-vs-MnTe}c. Based on the symmetry of the distribution of the magnon magnetic moment, we introduce a quantity $\lambda$ defined in Eq.~\eqref{eq:lambda}
which possesses the $g$-wave symmetry in both cases, regardless of the sign and magnitude of the anisotropy, see the bottom row of Fig.~\ref{fig:CrSb-vs-MnTe}.
Note that parameter $\lambda$ has the same analytical form $\lambda=-2\delta\varepsilon\,\Omega^-_{\vec{k}}$
for CrSb, as well as for MnTe.

\begin{figure}
    \centering
    \includegraphics[width=\columnwidth]{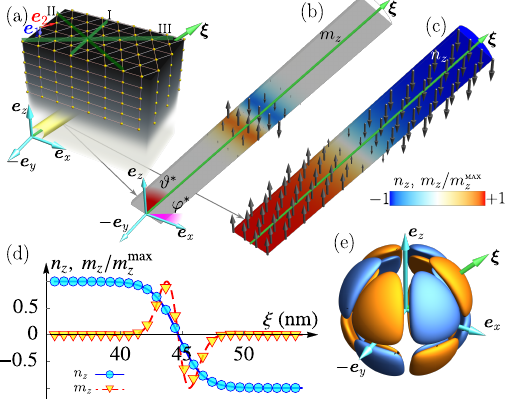}
    \caption{Domain wall magnetization: simulations vs theory. Panel (a) shows orientation of the domain wall direction $\xi$ relative to the crystal structure (yellow points denote Cr atoms). Vector $\vec{e}_\xi$ lies within plane III, its orientation is determined by the spherical angles $\vartheta^*$ and $\varphi^*=\pi/6$ which correspond to the maximal domain wall magnetization. Spin-lattice simulations were performed for Cr atoms located within a cylindrical tube of radius $3x_{\textsc{dw}}$ and oriented along $\vec{e}_\xi$. Cross-sections of the simulated tubes with radius $0.5x_{\textsc{dw}}$ are shown on panels (b) and (c) which demonstrate distributions of the magnetization and N{\'e}el vector, respectively. Arrows show the projection of $\vec{m}$ or $\vec{n}$ on $z$-axis. Panel (d) compares the profiles $n_z$ and $m_z$ obtained analytically (lines) and using the simulations (symbols). Panel (e) shows isosurfaces $m_z=\pm 10^{-3}$ for a hypothetical spherical domain wall $n_z=-\tanh[(\sqrt{x^2+y^2+z^2}-R)/x_{\textsc{dw}}]$, $n_y/n_x=\tan\phi_0$ with $R=7x_{\textsc{dw}}$, and $x_{\textsc{dw}} = 3$.}
    \label{fig:dw}
\end{figure}

\section{Magnetization of magnetic textures}

Applying the Taylor expansion to terms $\vec{m}_\nu(\vec{R}_{\vec{n}}+\vec{\delta R})$ and keeping terms  up to the 4th order in $\delta R_i$ we proceed to the continuum approximation of the Hamiltonian $\mathcal{H}$ by performing the replacement $\sum_{\vec{R}_{\vec{n}}}(\dots)\to\frac{1}{V_{\textsc{puc}}}\int(\dots)\dd\vec{r}$ with $V_{\textsc{puc}}=\sqrt{3}a_0^2c_0$ being the volume of the primitive magnetic unit cell. The continuum approximation of the altermagnetic part is 
\begin{equation}
    \label{eq:H-alt}\mathcal{H}_{\textsc{alt}}=\frac{\mathcal{B}_{\textsc{alt}}}{2}\int\left(\vec{m}_1\cdot\hat{\mathfrak{D}}\vec{m}_1-\vec{m}_2\cdot\hat{\mathfrak{D}}\vec{m}_2\right)\dd\vec{r},
\end{equation}
where $\mathcal{B}_{\textsc{alt}}=3a_0\,\delta\tilde{J}$ and $\hat{\mathfrak{D}}=\partial_{zy}(\partial_{yy}-3\partial_{xx})$. Note that the form of the differential operator $\hat{\mathfrak{D}}$ of the $g$-wave altermagnetic Hamiltonian \eqref{eq:H-alt} is very different to the one for $d$-wave altermagnets~\cite{Gomonay24a}. Note that the 4-th order of the Taylor expansion is the lowest order which leads to the nonvanishing $\mathcal{H}_{\textsc{alt}}$. The complete expression for $\mathcal{H}$ is provided in Eq.~\eqref{eq:H-cnt}.
Introducing the N{\'e}el order parameter $\vec{n}=(\vec{m}_1-\vec{m}_2)/2$ and the vector of the net magnetization $\vec{m}=(\vec{m}_1+\vec{m}_2)/2$, the magnetization can be shown to be  
\begin{equation}\label{eq:m}
    \vec{m}\approx\frac{1}{\gamma B_{ex}}\dot{\vec{n}}\times\vec{n}+\frac{1}{B_{ex}}\vec{n}\times\vec{B}\times\vec{n}-\mu_{\textsc{alt}}\vec{n}\times\hat{\mathfrak{D}}\vec{n}\times\vec{n},
\end{equation}
in the exchange approximation ($|\vec{m}|\ll1$), where $B_{ex}=4J_{\textsc{afm}}/\mu_s$ is the exchange field and $\mu_{\textsc{alt}}=\frac{3\sqrt{3}}{4}\frac{\delta\tilde{J}}{J_{\textsc{afm}}}a_0^3c_0$.

\subsection{Domain wall in CrSb}

The effective continuum Hamiltonian allows a domain wall solution for the easy-axial CrSb.  In terms of the continuous N{\'e}el vector represented in the angular parameterization $\vec{n}=\sin\theta(\vec{e}_x\cos\phi+\vec{e}_y\sin\phi)+\vec{e}_z\cos\theta$ the domain wall solution is $\theta=2\arctan e^{p\xi/x_{\textsc{dw}}}$ with $p=\pm1$ being the domain wall topological charge and $\phi=\phi_0=$const determines the domain wall chirality, e.g. $\phi_0=\pm\pi/2$ for Bloch domain walls of opposite chiralities. Here $\xi$ is coordinate along some arbitrary direction determined by a unit vector $\vec{e}_\xi=\sin\vartheta(\vec{e}_x\cos\varphi+\vec{e}_y\sin\varphi)+\vec{e}_z\cos\vartheta$, i.e. $\xi=\vec{r}\cdot\vec{e}_\xi$. The domain wall width $x_{\textsc{dw}}=(\ell_\perp^2\sin^2\vartheta+\ell_z^2\cos^2\vartheta)^{1/2}$ depends on the domain wall orientation. Applying \eqref{eq:m} we obtain the following magnetization of a static domain wall in the case $B=0$:
\begin{equation}
\begin{split}
\vec{m}=&\mu_{\textsc{alt}}\Upsilon(\vartheta,\varphi)\mathcal{R}(\xi/x_{\textsc{dw}})\\
&\times\left[\sinh(\xi/x_{\textsc{dw}})\left(\vec{e}_x\cos\phi_0+\vec{e}_y\sin\phi_0\right)+p\vec{e}_z\right].
\end{split}
\end{equation}
Function $\Upsilon(\vartheta,\varphi)=\tan^3\vartheta\sin\varphi(1-4\cos^2\varphi)/(\ell_\perp^2\tan^2\vartheta+\ell^2_z)^2$ determines the angular distribution of the magnetization magnitude concerning the domain wall orientation, and function $\mathcal{R}(x)=\tanh x(12-\cosh^2x)/\cosh^4x$  determines the magnetization distribution in the direction perpendicular to domain wall. The dependencies $n_z(\xi)$ and $m_z(\xi)$ for $p=1$ are shown in Fig.~\ref{fig:dw}(d). Function $\Upsilon(\vartheta,\varphi)$ has extrema for $\vartheta=\vartheta^*=\pm\arctan(\sqrt{3}\ell_z/\ell_\perp)\approx43^\circ$ and $\varphi=\varphi^*=\pm\frac{\pi}{2},\,\pm\frac{\pi}{6},\,\pi\pm\frac{\pi}{6}$. The extreme values $\varphi^*$ correspond to planes I, II, III shown in Figs.~\ref{fig:model}(a), \ref{fig:dw}(a). The extreme value is $\Upsilon(\vartheta^*,\varphi^*)=3\sqrt{3}/(16\ell_\perp^3\ell_z)$.  Using this and maximizing $\mathcal{R}(x)$ we estimate $m_z^{\text{max}}\approx\mu_{\textsc{alt}}/(\ell_{\perp}^3\ell_z)\approx 1.64\times 10^{-5}$. By introducing the magnetic moment density $\vec{M}=2\vec{m}\mu_s/V_{\textsc{puc}}$ we estimate $\mu_0M_z^{\text{max}}\approx14~\mu$T.

\section{Summary and conclusions} 
Based on simple Heisenberg models which capture the $g$-wave altermagnetic properties, we carried out the comparative analysis of the magnon spectra of easy-axial CrSb and easy-planar MnTe. We find that the altermagnetic splitting of the magnon branches of CrSb possesses $g$-wave symmetry and is the same as in the nonrelativistic limit (i.e., for vanishing anisotropy). In the case of MnTe, the easy-planar anisotropy qualitatively changes the magnon properties, and a generalization of the concept of altermagnetic splitting is required.  We propose a universal altemagnetic characteristic~\eqref{eq:lambda} that involves both the magnon energy and its magnetic moment and has $g$-wave symmetry in both cases, i.e. it is independent of the sign and magnitude of the anisotropy.


We expect that the described difference between the magnon properties in the easy-axial and easy-planar $g$-wave altermagnets also takes place for $d$-wave altermagnets. Namely, we expect the altermagnetic-induced magnon magnetic moment in easy-planar $d$-wave altermagnets, e.g. in NiF$_2$, whose distribution over the 1st Brillouin zone possesses $d$-wave symmetry for each magnon branch.

Similarly to $d$-wave altermagnets~\cite{Gomonay24a}, static domain walls in CrSb possess altermagnetically-induced magnetization which strongly depends on the domain wall orientation concerning the crystallographic axes. 

\appendix

\section{Exchange Hamiltonian}

Here we present the explicit form of the exchange part of the Hamiltonian. The antiferromagnetic interaction~($J_{\textsc{afm}}>0$) between layers~(001) has the following explicit form
\begin{equation}\label{eq:Hafm}
\begin{split}
	\mathcal{H}_{\textsc{afm}}=J_{\textsc{afm}}\sum\limits_{\vec{R}_{\vec{n}}}\vec{m}_1(\vec{R}_{\vec{n}})\cdot\bigl[&\vec{m}_2(\vec{R}_{\vec{n}}+c_0\vec{e}_z)\\
    +&\vec{m}_2(\vec{R}_{\vec{n}}-c_0\vec{e}_z)\bigr].
\end{split}
\end{equation}
The ferromagnetic exchange~($J_{\textsc{fm}}>0$) between the nearest neighbors within (001) layer is
\begin{equation}\label{eq:Hfm}
	\begin{split}
	&\mathcal{H}_{\textsc{fm}}=-J_{\textsc{fm}}\sum\limits_{\vec{R}_{\vec{n}}}\sum\limits_{\vec{\delta R}_\perp}\sum\limits_{\nu}\vec{m}_\nu(\vec{R}_{\vec{n}}+\vec{\zeta}_\nu)\\
    \cdot\bigl[\vec{m}_\nu&(\vec{R}_{\vec{n}}\!+\vec{\zeta}_\nu\!+\vec{\delta R}_\perp)\!+\vec{m}_\nu(\vec{R}_{\vec{n}}\!+\vec{\zeta}_\nu\!-\vec{\delta R}_\perp)\bigr],
	\end{split}
\end{equation}
where $\vec{\delta R}_\perp\in\{\vec{e}_1,\,\vec{e}_2,\,\vec{e}_1+\vec{e}_2\}$, $\vec{\zeta}_\nu = (\nu-1)c_0 \vec{e}_z$, and $\nu\in\{1,2\}$ enumerates sublatices. We present the altermagnetic Hamiltonian as a sum of symmetrical and asymmetrical parts $\mathcal{H}_{\textsc{alt}}=\mathcal{H}_{\textsc{alt},\textsc{s}}+\mathcal{H}_{\textsc{alt},\textsc{a}}$, where
\begin{subequations}\label{eq:Halt}
	\begin{align}
 	&\mathcal{H}_{\textsc{alt},\textsc{s}}=\frac{\tilde{J}}{2}\sum\limits_{\vec{R}_{\vec{n}}}\sum\limits_{\mathfrak{p}\in\{\textsc{i,ii,iii}\}}\!\!\sum\limits_{\nu}\vec{m}_\nu(\vec{R}_{\vec{n}}+\vec{\zeta}_\nu) \\
    &\cdot\Bigl[\vec{m}_\nu(\vec{R}_{\vec{n}}+\vec{\zeta}_\nu+\vec{\delta R}_{+}^{\mathfrak{p}})+\vec{m}_\nu(\vec{R}_{\vec{n}}+\vec{\zeta}_\nu-\vec{\delta R}_{+}^{\mathfrak{p}})\nonumber\\
    &+\vec{m}_\nu(\vec{R}_{\vec{n}}+\vec{\zeta}_\nu+\vec{\delta R}_{-}^{\mathfrak{p}})+\vec{m}_\nu(\vec{R}_{\vec{n}}+\vec{\zeta}_\nu-\vec{\delta R}_{-}^{\mathfrak{p}})\Bigr] \nonumber,\\
   	&\mathcal{H}_{\textsc{alt},\textsc{a}}=\frac{\delta\tilde{J}}{2}\sum\limits_{\vec{R}_{\vec{n}}}\sum\limits_{\mathfrak{p}\in\{\textsc{i,ii,iii}\}}\!\!\sum\limits_{\nu}(-1)^\nu\vec{m}_\nu(\vec{R}_{\vec{n}}+\vec{\zeta}_\nu)\\
    &\cdot\Bigl[\vec{m}_\nu(\vec{R}_{\vec{n}}+\vec{\zeta}_\nu+\vec{\delta R}_{+}^{\mathfrak{p}})+\vec{m}_\nu(\vec{R}_{\vec{n}}+\vec{\zeta}_\nu-\vec{\delta R}_{+}^{\mathfrak{p}})\nonumber\\
    &-\vec{m}_\nu(\vec{R}_{\vec{n}}+\vec{\zeta}_\nu+\vec{\delta R}_{-}^{\mathfrak{p}})-\vec{m}_\nu(\vec{R}_{\vec{n}}+\vec{\zeta}_\nu-\vec{\delta R}_{-}^{\mathfrak{p}})\Bigr]\nonumber.
	\end{align}
\end{subequations}

Table~\ref{tab:params-discr} lists the numerical values of the considered model's parameters.


\section{Magnon spectrum calculation}\label{app:magnon-spectrum}

\subsection{Dispersion relations}\label{app:magnon-disp}

\begin{figure*}
	\includegraphics[width=0.85\textwidth]{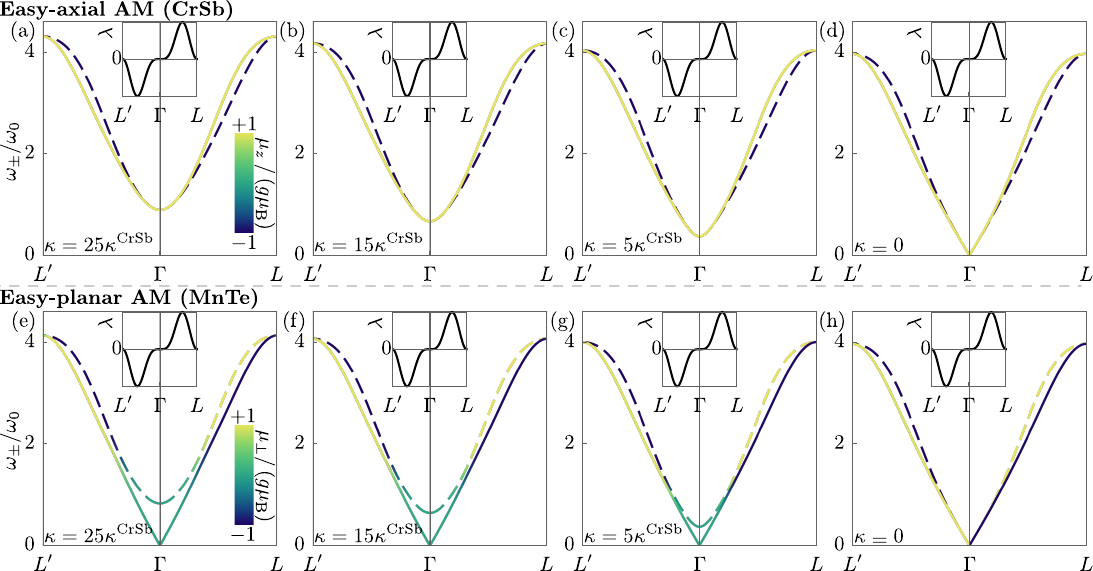}
	\caption{Comparison of the magnon spectra \eqref{eq:disp-CSb} for easy-axial AM (top row, CrSb) and \eqref{eq:disp-MnTe} for easy-planar AM (bottom row, MnTe) for different anisotropy strength. Color scheme corresponds to the amplitude of the magnon magnetic moment. The figures are built for the following values $\varepsilon=\varepsilon^\text{CrSb}$, $\delta\varepsilon=3\delta\varepsilon^\text{CrSb}$, $\eta=\eta^\text{CrSb}$, and $B=0$. Insets in the figures show the $g$-wave symmetry characteristic $\lambda$ for given anisotropy strength.}\label{fig:dispersion-CrSb-MnTe_vs_kappa}
\end{figure*}

Here we utilize the Fourier transforms on the periodic lattice 
\begin{equation}\label{eq:FT-discr}
	\begin{split}
		&\psi_\nu(\vec{R}_{\vec{n}})=\frac{1}{\sqrt{N}}\sum\limits_{\vec{k}\in1.\text{BZ}}\hat{\psi}_\nu(\vec{k})e^{i\vec{k}\cdot\vec{R}_{\vec{n}}},\\
		&\hat{\psi}_\nu(\vec{k})=\frac{1}{\sqrt{N}}\sum\limits_{\vec{R}_{\vec{n}}}\psi_\nu(\vec{R}_{\vec{n}})e^{-i\vec{k}\cdot\vec{R}_{\vec{n}}}	
	\end{split}
\end{equation}
supplemented with the completeness relation $\sum_{\vec{R}_{\vec{n}}}e^{i(\vec{k}-\vec{k}')\cdot \vec{R}_{\vec{n}}}=N\delta_{\vec{k},\vec{k}'}$. Here $N$ is the number of magnetic moments in one sublattice. Applying the Fourier transform~\eqref{eq:FT-discr} to linearized Eq.~\eqref{eq:LL-psi}, we obtain
\begin{equation}\label{eq:LL-psi-k}
    i\dot{\hat{\psi}}_\nu(\vec{k})=\frac{\gamma}{\mu_s}\frac{\partial\mathcal{H}^{(2)}}{\partial\hat{\psi}^*_\nu(\vec{k})},
\end{equation}
where $\mathcal{H}^{(2)}=\mathcal{H}_{\textsc{afm}}^{(2)}+\mathcal{H}_{\textsc{fm}}^{(2)}+\mathcal{H}_{\textsc{alt}}^{(2)}+\mathcal{H}_{\textsc{an}}^{(2)}+\mathcal{H}_{\textsc{z}}^{(2)}$ is harmonic (with respect to $\hat{\psi}_\nu$) part of Hamiltonian $\mathcal{H}$. The form of $\mathcal{H}^{(2)}$ depends on the ground state.

\underline{For CrSb} the ground state is $\vec{m}_1^0=(-1)^{\nu-1}\vec{e}_z$, with $\nu=1,2$. Without loss of generality one can take $\vec{T}_\nu=[(-1)^{\nu-1}\vec{e}_x+i\vec{e}_y]/2$ \footnote{In general case, $\vec{T}_\nu=e^{i\Phi_\nu}[(-1)^{\nu-1}\vec{e}_x+i\vec{e}_y]/2$, where $\Phi_\nu$ are constant phases. One can always get rid of the constant phase factors by the change of variables $\tilde{\psi}_\nu=e^{i\Phi_\nu}\psi_\nu$ which keeps the equation \eqref{eq:LL-psi} unchanged.}. Now using \eqref{eq:Tyabl}, form of Hamiltonian $\mathcal{H}$ and Fourier transform \eqref{eq:FT-discr}, we derive the Hamiltonian components in the harmonic approximation
\begin{equation}\label{eq:H-harm-k-CSb}
    \begin{split}
    &\mathcal{H}_{\textsc{afm}}^{(2)}=2J_{\textsc{afm}}\!\!\sum\limits_{\vec{k}\in\text{1.BZ}}\!\!\biggl\{\hat{\Xi}^{+}_{\vec{k}}-\cos(k_zc_0)\biggl[\hat{\psi}_1(\vec{k})\hat{\psi}_2(-\vec{k})\\
    &+\hat{\psi}^*_1(\vec{k})\hat{\psi}^*_2(-\vec{k})\biggr]\biggr\},\\
    &\mathcal{H}_{\textsc{an}}^{(2)}=2K\!\!\sum\limits_{\vec{k}\in\text{1.BZ}}\!\!\hat{\Xi}^{+}_{\vec{k}},\quad\mathcal{H}_{\textsc{z}}^{(2)}=\mu_s B\!\!\sum\limits_{\vec{k}\in\text{1.BZ}}\!\!\hat{\Xi}^{-}_{\vec{k}},\\
    &\mathcal{H}_{\textsc{fm}}^{(2)}+\mathcal{H}_{\textsc{alt}}^{(2)}=4\!\!\sum\limits_{\vec{k}\in\text{1.BZ}}\!\!\left[\left(2J_{\textsc{fm}}\Omega^{\textsc{fm}}_{\vec{k}}-\tilde{J}\Omega^{+}_{\vec{k}}\right)\hat{\Xi}^{+}_{\vec{k}}+\delta\tilde{J}\Omega^{-}_{\vec{k}}\hat{\Xi}^{-}_{\vec{k}}\right],
    \end{split}
\end{equation}
where $\hat{\Xi}^{\pm}_{\vec{k}}=|\hat{\psi}_1(\vec{k})|^2\pm|\hat{\psi}_2(\vec{k})|^2$. Thus, the total harmonic Hamiltonian can be written in the form 
\begin{equation}\label{eq:H-harm-CrSb}
    \begin{split}
        &\mathcal{H}^{(2)}=\omega_0\frac{\mu_s}{\gamma}\!\!\sum\limits_{\vec{k}\in\text{1.BZ}}\!\!\Bigl\{\mathcal{A}_{\vec{k}}\hat{\Xi}^{+}_{\vec{k}}+\mathcal{B}_{\vec{k}}\hat{\Xi}^{-}_{\vec{k}}\\
        &-\cos(k_zc_0)\left[\hat{\psi}_1(\vec{k})\hat{\psi}_2(-\vec{k})+\hat{\psi}^*_1(\vec{k})\hat{\psi}^*_2(-\vec{k})\right]\Bigr\},
    \end{split}
\end{equation}
where $\mathcal{A}_{\vec{k}}=1+\eta\Omega_{\vec{k}}^{\textsc{fm}}-\varepsilon\Omega_{\vec{k}}^{+}+\kappa=F_{\vec{k}}+\kappa$, and $\mathcal{B}_{\vec{k}}=\frac{\gamma B}{\omega_0}+\delta\varepsilon\Omega^{-}_{\vec{k}}$. With Hamiltonian \eqref{eq:H-harm-CrSb}, the couple of equations \eqref{eq:LL-psi-k} together with their complex conjugated counterparts form a closed set of four equations
\begin{align}\label{eq:Psi-CrSb}
    &i\dot{\hat{\vec{\Psi}}}=\mathbb{M}\hat{\vec{\Psi}},\\\nonumber
    &\frac{\mathbb{M}}{\omega_0}=\begin{bmatrix}
    \mathcal{A}_{\vec{k}}+\mathcal{B}_{\vec{k}} & 0 & 0 & -\cos(k_zc_0) \\
    0 & \mathcal{A}_{\vec{k}}-\mathcal{B}_{\vec{k}} & -\cos(k_zc_0) & 0 \\
    0 & \cos(k_zc_0) & -(\mathcal{A}_{\vec{k}}+\mathcal{B}_{\vec{k}}) & 0 \\
    \cos(k_zc_0) & 0 & 0 & -(\mathcal{A}_{\vec{k}}-\mathcal{B}_{\vec{k}})
    \end{bmatrix}
\end{align}
for each given $\vec{k}$. Here $\hat{\vec{\Psi}}=[\hat{\psi}_1(\vec{k}),\hat{\psi}_2(\vec{k}),\hat{\psi}^*_1(-\vec{k}),\hat{\psi}^*_2(-\vec{k})]^{\textsc{t}}$ and we used that $\mathcal{A}_{\vec{k}}=\mathcal{A}_{-\vec{k}}$, $\mathcal{B}_{\vec{k}}=\mathcal{B}_{-\vec{k}}$. The set \eqref{eq:Psi-CrSb} has solution $\hat{\vec{\Psi}}=\hat{\vec{\Psi}}_0e^{-i\omega t}$, which is nontrivial ($\hat{\vec{\Psi}}_0\ne\vec{0}$) when $\omega$ is an eigenvalue of $\mathbb{M}$. This gives the dispersion relation \eqref{eq:disp-CSb}.

\underline{For MnTe}, the ground state is $\vec{m}_\nu^0=(-1)^{\nu-1}\sin\theta_0(\vec{e}_x\cos\phi_0+\vec{e}_y\sin\phi_0)+\vec{e}_z\cos\theta_0$ and consequently $\vec{T}_\nu=\frac{(-1)^{\nu-1}}{2}(\cos\theta_0\cos\phi_0-i\sin\phi_0)\vec{e}_x+\frac{(-1)^{\nu-1}}{2}(\cos\theta_0\sin\phi_0+i\cos\phi_0)\vec{e}_y-\frac{1}{2}\sin\theta_0\vec{e}_z.$ The harmonic part of the Hamiltonian has the following contributions
\begin{widetext}
\begin{equation}
    \begin{split}
        &\mathcal{H}_{\textsc{afm}}^{(2)}=J_{\textsc{afm}}\!\!\sum\limits_{\vec{k}\in\text{1.BZ}}\!\!\left\{-\cos2\theta_0\hat{\Xi}^{+}_{\vec{k}}+2\cos(k_zc_0)\left[\sin^2\theta_0\hat{\psi}_1(\vec{k})\hat{\psi}_2(-\vec{k})-\cos^2\theta_0\hat{\psi}_1(\vec{k})\hat{\psi}^*_2(\vec{k})\right]+\text{c.c.}\right\},\\
        &\mathcal{H}_{\textsc{an}}^{(2)}\!=\!|K|\!\!\!\sum\limits_{\vec{k}\in\text{1.BZ}}\sum\limits_{\nu}\left\{\frac{\sin^2\theta_0}{2}\hat{\psi}_\nu(\vec{k})\hat{\psi}_\nu(-\vec{k})+\frac{1-3\cos^2\theta_0}{2}|\hat{\psi}_\nu|^2+\text{c.c.}\right\},\quad\mathcal{H}_{\textsc{z}}^{(2)}=B\mu_s\cos\theta_0\sum\limits_{\vec{k}\in\text{1.BZ}}\!\hat{\Xi}^{+}_{\vec{k}}.
    \end{split}
\end{equation} 
The expressions for $\mathcal{H}_{\textsc{alt}}^{(2)}$ and $\mathcal{H}_{\textsc{fm}}^{(2)}$ are the same as in \eqref{eq:H-harm-k-CSb}. For each $\vec{k}$, equations \eqref{eq:LL-psi-k} together with their complex conjugated counterparts form the set \eqref{eq:Psi-CrSb} with matrix
\begin{equation}\label{eq:Psi-MnTe}
\mathbb{M}\!=\!\omega_0\!\begin{bmatrix}
        F_{\vec{k}}+\frac{\kappa}{2}\sin^2\theta_0-\delta\varepsilon\Omega^-_{\vec{k}} & -\cos^2\theta_0\cos(k_zc_0) & \frac{\kappa}{2}\sin^2\theta_0 & \sin^2\theta_0\cos(k_zc_0) \\
        -\cos^2\theta_0\cos(k_zc_0) & F_{\vec{k}}+\frac{\kappa}{2}\sin^2\theta_0+\delta\varepsilon\Omega^-_{\vec{k}} & \sin^2\theta_0\cos(k_zc_0) & \frac{\kappa}{2}\sin^2\theta_0 \\
        -\frac{\kappa}{2}\sin^2\theta_0 & -\sin^2\theta_0\cos(k_zc_0) & -(F_{\vec{k}}+\frac{\kappa}{2}\sin^2\theta_0-\delta\varepsilon\Omega^-_{\vec{k}}) & \cos^2\theta_0\cos(k_zc_0) \\
        -\sin^2\theta_0\cos(k_zc_0) & -\frac{\kappa}{2}\sin^2\theta_0 & \cos^2\theta_0\cos(k_zc_0) & -(F_{\vec{k}}+\frac{\kappa}{2}\sin^2\theta_0+\delta\varepsilon\Omega^-_{\vec{k}})
    \end{bmatrix},
\end{equation}    
\end{widetext}
whose eigenvalues determine the dispersion relation~\eqref{eq:disp-MnTe}. Here $\kappa=|K|/J_{\textsc{afm}}$ and definitions of the constants $\omega_0$, $\eta$, $\varepsilon$, $\delta\varepsilon$ are the same as previously.

Solving the eigenvalue problem we compute also the eigenvector $\hat{\vec{\Psi}}_0(\vec{k})=[\hat{\psi}_{01}(\vec{k}),\hat{\psi}_{02}(\vec{k}),\hat{\psi}^*_{01}(-\vec{k}),\hat{\psi}^*_{02}(-\vec{k})]^{\textsc{t}}$ which enables us to reconstruct dynamics of a magnetic moment sitting in the site $\vec{R}_{\vec{n}}$ when a spin wave with the wave-vector $\vec{k}$ is excited:
\begin{equation}\label{eq:dm-sw}
\begin{split}
    \vec{m}_\nu\approx\vec{m}_\nu^0+\sqrt{\frac{2}{N}}&\,\mathrm{Re}\biggl[e^{i(\vec{k}\cdot\vec{R}_{\vec{n}}-\omega_{\vec{k}}t+\varphi_0)}\\
    &\times\left(\vec{T}_\nu\hat{\psi}_{0\nu}(\vec{k})+\vec{T}^*_\nu\hat{\psi}^*_{0\nu}(-\vec{k})\right)\biggr].
\end{split}
\end{equation}
Here $\varphi_0$ is an arbitrary phase. Examples of the reconstructed dynamics are shown in Fig.~\ref{fig:dispersion-CrSb-MnTe}(e,j) and Fig.~\ref{fig:precess-MnTe}.

\subsection{Magnetic moment of a magnon in MnTe}
As was shown in the previous subsection, the field dependence of the eigenfrequencies has a general form $\omega_\nu=\omega_\nu(B^2)$ if the magnetic field is applied perpendicularly to the ground state. According to \eqref{eq:mu}, this leads to the vanishing of the component of the magnetic moment perpendicular to the ground state. To compute magnetic moment in the direction of the ground state of MnTe, we introduce an auxiliary easy-axial anisotropy with the easy axis lying within the easy plane
\begin{equation}\label{eq:Han-aux}
    \tilde{\mathcal{H}}_{\textsc{an}}=-\tilde{K}\sum_{\vec{R}_{\vec{n}}}\left[m_{1x}^2(\vec{R}_{\vec{n}})+m_{2x}^2(\vec{R}_{\vec{n}}+c_0\vec{e}_z)\right].
\end{equation}
Due to \eqref{eq:Han-aux}, the orientation of the ground state within the easy plane is fixed in $x$-direction and application of a small enough magnetic field in this direction does not lead to the spin-flop reorientation. Now, we apply the technique described in Sec.~\ref{app:magnon-disp} to the MnTe Hamiltonian with the additional term \eqref{eq:Han-aux} and with the Zeeman term in form $\mathcal{H}_{\textsc{z}}=-B\mu_s\sum_{\vec{R}_{\vec{n}}}\left[m_{1x}(\vec{R}_{\vec{n}})+m_{2x}(\vec{R}_{\vec{n}}+c_0\vec{e}_z)\right]$ and obtain the following dispersion relation for magnons excited on the top of the ground state $\vec{m}_\nu^0=(-1)^{\nu+1}\vec{e}_x$
\begin{align}\label{eq:omega-MnTe-inplane}
    \omega_\pm=&\omega_0\sqrt{\tilde{\mathcal{A}}_{\vec{k}}^2+\tilde{\mathcal{B}}_{\vec{k}}^2-\cos^2(k_zc_0)-\frac{\kappa^2}{4}\pm2\varpi_{\vec{k}}},\\\nonumber
    &\varpi_{\vec{k}}=\sqrt{\tilde{\mathcal{A}}_{\vec{k}}^2\tilde{\mathcal{B}}_{\vec{k}}^2-\cos^2(k_zc_0)(\tilde{\mathcal{B}}_{\vec{k}}^2-\kappa^2/4)}.
\end{align}
Here $\tilde{\mathcal{A}}_{\vec{k}}=F_{\vec{k}}+\frac{\kappa}{2}+\tilde{\kappa}$ with $\tilde{\kappa}=\tilde{K}/J_{\textsc{afm}}$, and $\tilde{\mathcal{B}}_{\vec{k}}=\delta\varepsilon\,\Omega^-_{\vec{k}}+\frac{\gamma B}{\omega_0}$. Note that dispersions \eqref{eq:omega-MnTe-inplane} and \eqref{eq:disp-MnTe} coincide for the case $B=0$ and $\tilde{\kappa}=0$. Using Eq.~\eqref{eq:mu} and applying the limit $\tilde{\kappa}\to0$ we obtain
\begin{align}\label{eq:mu-exact}
    \nonumber&(\mu^{\text{MnTe}}_\pm)_\perp\!\!=\! \frac{-g\mu_{\textsc{b}}\delta\varepsilon\,\Omega^-_{\vec{k}}\left[1\pm\frac{(F_{\vec{k}}+\frac{\kappa}{2})^2-\cos^2(k_zc_0)}{\varpi_{\vec{k}}^0}\right]}{\sqrt{F_{\vec{k}}(F_{\vec{k}}+\frac{\kappa}{2})+(\delta\varepsilon\Omega^{-}_{\vec{k}})^2-\cos^2(k_zc_0)\pm\varpi_{\vec{k}}^0}},\\
    &\varpi_{\vec{k}}^0\!=\!\sqrt{\left(F_{\vec{k}}+\frac{\kappa}{2}\right)^2\!\!(\delta\varepsilon\Omega^{-}_{\vec{k}})^2-\cos^2(k_zc_0)\left[(\delta\varepsilon\Omega^{-}_{\vec{k}})^2-\frac{\kappa^2}{4}\right]}
    \end{align}
meaning that the total moment is $\vec{\mu}^{\text{MnTe}}_\pm=(\mu^{\text{MnTe}}_\pm)_\perp\vec{e}_x$. In general case, magnons with the same $\vec{k}$ belonging to different branches have magnetic moments of diferent amplitudes, i.e. $|(\mu^{\text{MnTe}}_+)_\perp|\ne|(\mu^{\text{MnTe}}_-)_\perp|$. However, for the parameters of MnTe (see Table~\ref{tab:params-discr}) one has $(\mu^{\text{MnTe}}_+)_\perp\approx-(\mu^{\text{MnTe}}_-)_\perp$, see Fig.~\ref{fig:magnetization-MnTe}. The straightforward calculation of parameter $\lambda$ with the use of \eqref{eq:omega-MnTe-inplane} and \eqref{eq:mu-exact} results in $\lambda=-2\delta\varepsilon\,\Omega^-_{\vec{k}}$.

\section{Continuous approximation}

To obtain the continuous approximation of the Hamiltonian $\mathcal{H}$ we utilize the Taylor expansion up to the 4th order $\vec{m}_\nu(\vec{R}_{\vec{n}}+\vec{\delta R})\approx\vec{m}_\nu(\vec{R}_{\vec{n}})+\delta R_i\partial_i\vec{m}_\nu(\vec{R}_{\vec{n}})+\frac{1}{2}\delta R_i\delta R_j\partial_{ij}\vec{m}_\nu(\vec{R}_{\vec{n}})+\frac{1}{3!}\delta R_i\delta R_j\delta R_k\partial_{ijk}\vec{m}_\nu(\vec{R}_{\vec{n}})+\frac{1}{4!}\delta R_i\delta R_j\delta R_k\delta R_l\partial_{ijkl}\vec{m}_\nu(\vec{R}_{\vec{n}})$
and perform the replacement $\sum_{\vec{R}_{\vec{n}}}(\dots)\to\frac{1}{V_{\textsc{puc}}}\int(\dots)\dd\vec{r}$ with $V_{\textsc{puc}}=\sqrt{3}a_0^2c_0$ being the volume of the primitive magnetic unit cell. Finally,
\begin{widetext}
\begin{equation}\label{eq:H-cnt}
\begin{split}
    \mathcal{H}=&\int\Bigl[\frac12B_{ex}M_s\,\vec{m}_1\cdot\vec{m}_2-A_{\textsc{afm}}\partial_z\vec{m}_1\cdot\partial_z\vec{m}_2+\frac12A_{\textsc{fm}}^\perp\partial_\alpha\vec{m}_\nu\cdot\partial_\alpha\vec{m}_\nu+\frac12A_{\textsc{fm}}^z\partial_z\vec{m}_\nu\cdot\partial_z\vec{m}_\nu-\frac12B_{an}M_s(m_{1z}^2+m_{2z}^2)\\
    &-M_s\vec{B}\cdot(\vec{m}_{1}+\vec{m}_{2})+\frac12\mathcal{B}_{\textsc{alt}}\left(\vec{m}_1\cdot\hat{\mathfrak{D}}\vec{m}_1-\vec{m}_2\cdot\hat{\mathfrak{D}}\vec{m}_2\right)+B_{\textsc{afm}}\partial_{zz}\vec{m}_1\cdot\partial_{zz}\vec{m}_2+\frac12B_{ij}\partial_{ii}\vec{m}_\nu\cdot\partial_{jj}\vec{m}_\nu\Bigr]\dd\vec{r}
    \end{split}
\end{equation}
\end{widetext}
Here indices $\alpha\in\{x,y\}$ and $i,j\in\{x,y,z\}$ numerate coordinates, $\nu=1,2$ numerates the sublattices and the summation over the repeating indices is assumed. We introduced the following parameters: $M_s=\mu_s/V_{\textsc{puc}}$ is the saturation magnetization of one sublattice, $A_{\textsc{afm}}=J_{\textsc{afm}}c_0/(\sqrt{3}a_0^2)$ is antiferromagnetic stiffness, $A_{\textsc{fm}}^\perp=\sqrt{3}(J_{\textsc{fm}}-3\tilde{J})/c_0$ is ferromagnetic stiffness in $xy$-plane, $A_{\textsc{fm}}^z=-8\sqrt{3}\tilde{J}c_0/a_0^2$ is stiffness in $z$-direction, $B_{an}=2K/\mu_s$ is the anisotropy field, $B_{\textsc{afm}}=J_{\textsc{afm}}c_0^3/(12\sqrt{3}a_0^2)$, and tensor $B_{ij}$
has components $B_{xx}=B_{yy}=B_{xy}=B_{yx}=-\frac{\sqrt{3}}{16}\frac{a_0^2}{c_0}(J_{\textsc{fm}}-9\tilde{J})$, $B_{xz}=B_{zx}=B_{yz}=B_{zy}=3\sqrt{3}c_0\tilde{J}$, $B_{zz}=\frac{8}{\sqrt{3}}\frac{c_0^3}{a_0^2}\tilde{J}$. 

The introduced constants determine typical lengthscales in $xy$-plane $\ell_\perp=\sqrt{A_{\textsc{fm}}^\perp/(M_sB_{an})}$ and in $z$-direction $\ell_z=\sqrt{(A_{\textsc{fm}}^z+A_{\textsc{afm}})/(M_sB_{an})}$. Typical timescale is determined by the frequency of uniform ferromagnetic resonance $\omega_{\textsc{afmr}}=\gamma B_{sf}$ with $B_{sf}=\sqrt{B_{ex}B_{an}}$ being the spin-flop field. The parameters $\ell_\perp$, $\ell_z$, $\omega_{\textsc{afmr}}$, and $B_{sf}$ coincide with the same parameters in the main text determined via constants of the discrete model.

%

\end{document}